\documentclass[twocolumn,twoside,slac]{revtex4}
\usepackage{graphicx}
\usepackage{fancyhdr}
\usepackage{longtable}
\begin{document}

\title{Performance comparison between iSCSI and other hardware and software
solutions}

%

\author{Mathias Gug}
\affiliation{CERN, Geneva, Switzerland}

\begin{abstract}
We report on our investigations on some technologies that can be used
to build disk servers and networks of disk servers using commodity
hardware and software solutions. It focuses on the performance that
can be achieved by these systems and gives measured figures for different
configurations.

It is divided into two parts : iSCSI and other technologies and hardware and software RAID solutions.

The first part studies different technologies that can be used by
clients to access disk servers using a gigabit ethernet network. It
covers block access technologies (iSCSI, hyperSCSI, ENBD). Experimental
figures are given for different numbers of clients and servers. 

The second part compares a system based on 3ware hardware RAID controllers,
a system using linux software RAID and IDE cards and a system mixing
both hardware RAID and software RAID. Performance measurements for
reading and writing are given for different RAID levels.
\end{abstract}

\maketitle

\thispagestyle{fancy}


\section{iSCSI, HyperSCSI and ENBD technologies}

\subsection{What is iSCSI ?}

\emph{iSCSI} is a protocol designed to transport SCSI commands over
a TCP/IP network.\\
\emph{iSCSI} can be used as a building block for network storage using
existing IP infrastructure in a LAN/WAN environment. It can connect
different types of block-oriented storage devices to servers.\\
\emph{iSCSI} was initially standardized by ANSI T10 and further developed
by the IP Storage working group of the IETF \cite{ietf-ips}, which
will publish soon an RFC. Many vendors in the storage industry as
well as research projects are currently working on the implementation
of the iSCSI protocol.

\begin{quotation}
\char`\"{}The Small Computer Systems Interface (SCSI) is a popular
family of protocols for communicating with I/O devices, especially
storage devices. SCSI is a client-server architecture. Clients of
a SCSI interface are called \char`\"{}initiators\char`\"{}. Initiators
issue SCSI \char`\"{}commands\char`\"{} to request services from components,
logical units, of a server known as a \char`\"{}target\char`\"{}.
A \char`\"{}SCSI transport\char`\"{} maps the client-server SCSI protocol
to a specific interconnect. Initiators are one endpoint of a SCSI
transport and targets are the other endpoint.

The iSCSI protocol describes a means of transporting of the SCSI packets
over TCP/IP, providing for an interoperable solution which can take
advantage of existing Internet infrastructure, Internet management
facilities and address distance limitations.\char`\"{}

draft-ietf-ips-iscsi-20
\end{quotation}

\subsection{What is HyperSCSI ?}

\emph{HyperSCSI} is a protocol that sends SCSI commands using raw
Ethernet packets instead of the TCP/IP packets used for \emph{iSCSI}.
Thus, it bypasses the TCP/IP stack of the OS and does not suffer from
the shortcomings of TCP/IP.

\emph{HyperSCSI} focuses on turning ethernet into a usable storage
infrastructure by adding missing components such as flow control,
segmentation, reassembly, encryption, access control lists and security.
It can be used to connect different type of storage, such as SCSI,
IDE and USB devices.

\emph{HyperSCSI} is developed by the \emph{Modular Connected Storage
Architecture} group in the Network Storage Technology Division of
the Data Storage Institute from the Agency for Science, Technology
and Research of Singapore \cite{hyperSCSI}.

\subsection{What is Enhanced Network Block Device (ENBD)?}

ENBD is a linux kernel module coupled with a user space daemon that
sends block requests from a linux client to a linux server using a
TCP/IP connection. It uses multichannel communications and implements
internal failover and automatic balancing between the channels. It
supports encryption and authentication.\\
This block access technology is only useful with a linux kernel because
of the linux specific block request format.\\
It is developed by the linux community \cite{enbd} under a GPL license.

\section{Test configuration}

\subsection{Hardware and software configuration}

The following hardware was used to perform the tests :

\begin{itemize}
\item \emph{Test2} : \\
Dual Pentium 3 - 1 Ghz\\
3Com Gigabit Ethernet card based on BROADCOM BCM 5700 chipset\\
1 Western Digital WD1800JB 180 Gbytes\\
3ware RAID controller 7000-series
\item \emph{Test11} :\\
Dual Pentium 4 - 2.4 Ghz (HyperThreading enabled)\\
6 Western Digital WD1800JB 180 Gbytes\\
3ware RAID controllers 7000-series or Promise Ultra133 IDE controllers\\
3Com Gigabit Ethernet card based on BROADCOM BCM 5700 chipset
\item \emph{Test13} :\\
Dual AMD MP 2200+\\
6 Western Digital WD1800JB 180 Gbytes\\
3ware RAID controllers 7000-series or Promise Ultra133 IDE controllers\\
3Com Gigabit Ethernet card based on BROADCOM BCM 5700 chipset
\item iSCSI server : Eurologic eLANtra iCS2100 IP-SAN storage appliance
- v1.0 \cite{eurologic}\\
3 SCSI drives
\end{itemize}
All the machines have a Redhat 7.3 based distribution, with kernel
2.4.19 or 2.4.20.\\
The following optimizations were made to improve the performance :

{\tt sysctl~-w~vm.min-readahead=127}

{\tt sysctl~-w~vm.max-readahead=256}

{\tt sysctl~-w~vm.bdflush~=}

{\tt '2~500~0~0~500~1000~60~20~0'}

{\tt elvtune~-r~512~-w~1024~/dev/hd\{a,c,e,g,i,k\}}

\subsection{Benchmarks and monitor}

Two benchmarks were used to measure the IO bandwidth and the CPU load
on the machines :

\begin{itemize}
\item \emph{bonnie++} : v 1.03 \cite{bonnie}\\
This benchmark measures the performance of harddrives and filesystems.
It aims at simulating a database like access pattern.\\
We are interested in two results : '\emph{sequential output block}'
and '\emph{sequential input block}'.\\
Bonnie++ uses a filesize of 9GBytes with a chunksize of 8KBytes. Bonnie++
reports the CPU load for each test. However, we found that the reported
CPU load war incorrect. So we used a standard monitoring tool (vmstat)
to measure the CPU Load during bonnie++ runs instead.
\item \emph{seqent\_random\_io64 :}\\
In this benchmark, we were interested in three results : 

\begin{itemize}
\item \emph{'write'} performance : bandwidth measured for writing a file
of 5 Gbytes, with a blocksize of 1.5 Mbytes.
\item \emph{'sequential reading'} performance : bandwidth measured for sequential
reading of a file of 5 Gbytes with a blocksize of 1.5 Mbytes.
\item \emph{'random reading'} performance : bandwidth measured for random
reads within a file of 5 Gbytes with a blocksize of 1.5 Mbytes.
\end{itemize}
This benchmark is a custom program used at CERN to evaluate the performance
of disk servers. It simulates an access pattern used by CERN applications.

\end{itemize}
\emph{vmstat} has been used to monitor the CPU load on each machine.

\section{Performance of iSCSI, HyperSCSI and ENBD}

\subsection{iSCSI performance for different software initiators}

The server was the Eurologic iCS2100 IP-SAN storage appliance \cite{eurologic}.
The client was \emph{test13}, with kernel 2.4.19smp.

Two software initiators were used to connect to the iSCSI server :
ibmiscsi \cite{ibmiscsi} and linux-iscsi \cite{linux-iscsi}. We
used two versions of linux-iscsi : 2.1.2.9, implementing version 0.8
of the iscsi draft, and 3.1.0.6, implementing version 0.16 of the
iscsi draft.

The results are given in the table below :

\begin{longtable}{|c|c|c|c|c|c|}
\hline 
&
seq output&
seq input&
write&
seq&
random \\
&block&
block& & read&read \\
\hline 
ibmiscsi&
42 MB/s&
60 &
38 &
58 &
39 \\
1.2.2&
37 \% CPU&
82 \% &
36 \%&
84 \% &
66 \% \\
\hline 
linux-iscsi&
62 MB/s&
62 &
60 &
60 &
42 \\
2.1.2.9&
43 \%&
78 \% &
44 \% &
82 \% &
42 \% \\
\hline 
linux-iscsi&
64 MB/s&
57 &
59 &
58 &
38 \\
3.1.0.6&
61 \%&
99 \% &
62 \% &
99 \% &
78 \% \\
\hline
\end{longtable}

\emph{Comments} :

\begin{itemize}
\item The maximum measured bandwidth of 60 MBytes/s corresponds to the maximum
throughput that the disks in the server can deliver : there were only
three SCSI drives in the server, each delivering a maximum throughput
of around 20 Mbytes/s.
\item In a storage infrastructure, the CPU load on the client should be
taken into account. In order to increase the overall performance of
the client, offloading engine cards were also considered. Unfortunately,
no driver for our Linux platform was available at that time.
\item During the performance measurements which represent several days of
continuous disk access, no crashes of either the client or the server
were observed.
\end{itemize}

\subsection{HyperSCSI and ENBD}

The measurements were made using block client (\emph{test11}) connected
to either one or three block servers. The kernel used was 2.4.19.

The results are given in the table below :

\begin{longtable}{|c|c|c|c|c|c|}
\hline 
&
seq output&
seq input&
write&
seq&
random \\
&block&
block& & read&read \\
\hline 
HyperSCSI&
46 MB/s&
43 &
46 &
44 &
21 \\
one server&
60 \%CPU&
81 \% &
61 \%&
79 \% &
50 \% \\
\hline 
ENBD &
27 &
28 &
19 &
28 &
19 \\
one server&
22 \%&
42 \% &
21 \%&
46 \% &
30 \% \\
\hline 
ENBD&
74 &
44 &
43 &
51 &
30 \\
three server&
46 \%&
18 \% &
45 \%&
19 \% &
22 \% \\
\hline
\end{longtable}

\emph{Comments :}

\begin{itemize}
\item Compared with local disk access, HyperSCSI is able to access the remote
drive at its maximum throughput.
\item HyperSCSI performs poorly if there are two or more connections to
different servers. More generally, HyperSCSI lacks stability and works
only on uniprocessor kernels 2.4.19.
\item ENBD is able to handle several connections to different servers and
integrates well within a software RAID array.
\end{itemize}

\section{Local performance tests using hardware and software solutions.}

\subsection{Software RAID performance}

This series of benchmarks was done on \emph{test11} with 6 harddrives,
using kernel 2.4.20. The hardware used was only composed of IDE drives
and IDE controllers. No hardware accelerator cards were used.

The results are given in the table below :

\begin{longtable}{|c|c|c|c|c|c|}
\hline 
&
seq output&
seq input&
write&
seq&
random \\
&block&
block& & read&read \\
\hline 
RAID 0&
170 MB/s&
166&
166 &
170 &
61 \\
&
23 \%CPU&
20 \% &
25 \%&
25 \% &
8 \% \\
\hline 
RAID 5&
85 &
131 &
79 &
140 &
62 \\
&
30 \%&
22 \% &
27 \%&
13 \% &
7 \%\\
\hline
\end{longtable}

\subsection{Software RAID and hardware RAID}

This series of benchmarks were done on \emph{test11} with 4 harddrives,
using kernel 2.4.20.

The results are given in the table below :

\begin{longtable}{|c|c|c|c|c|c|}
\hline 
&
seq output&
seq input&
write&
seq&
random \\
&block&
block& & read&read \\
\hline 
Software &
60 MB/s&
73 &
59 &
75 &
45 \\
RAID5&
30 \%CPU&
22 \% &
24 \%&
12 \% &
7 \% \\
\hline 
Hardware &
47 &
59 &
49 &
61 &
38 \\
RAID5&
6 \%&
10 \% &
2 \%&
6 \% &
4 \% \\
\hline 
Software&&&&&\\
RAID0 +&
84 &
54 &
80 &
56 &
40 \\
Hardware&
23 \%&
10 \% &
9 \%&
6 \% &
5 \% \\
RAID1&&&&&\\
\hline
\end{longtable}

\emph{Comments :}

SoftwareRAID delivers more bandwidth than HardwareRAID, but at a higher
CPU cost.

\section{Acknowledgments}

I would like to thank all the people from the IT/ADC group at CERN
for helping me in this study and particularly Markus Schulz, Arie
Van Praag, Remi Tordeux, Oscar Ponce Cruz, Jan Iven, Peter Kelemen
and Emanuele Leonardi for their support, comments and ideas.

I would also like to thank Eurologic for their iSCSI storage appliance
that was used during the evaluation of the iSCSI initiators.

I am currently part of the {}``Volontaire Internatinal'' program,
which is funded by the French Foreign Ministry.

\end{document}